\documentclass[%
reprint,
superscriptaddress,
 amsmath,amssymb,
 aps,
prl,
]{revtex4-1}
\usepackage{graphicx}
\usepackage{dcolumn}
\usepackage{bm}

\begin{document}

\title{Direct Visualization of Trimerized States in 1T$'$-TaTe$_{2}$}

\author{Ismail El Baggari}
 \affiliation{Department of Physics, Cornell University, Ithaca, New York 14853, USA}
\author{Nikhil Sivadas}
 \affiliation{School of Applied and Engineering Physics, Cornell University, Ithaca, New York 14853, USA}
\author{Gregory M. Stiehl}
 \affiliation{Department of Physics, Cornell University, Ithaca, New York 14853, USA}
 \author{Jacob Waelder}
 \affiliation{Platform for the Accelerated Realization, Analysis and Discovery of Interface Materials (PARADIM), Cornell University, Ithaca, New York 14853, USA}
 \author{Daniel C. Ralph}
\affiliation{Department of Physics, Cornell University, Ithaca, New York 14853, USA}  
\affiliation{Kavli Institute at Cornell for Nanoscale Science, Cornell University, Ithaca, New York 14853, USA}
\author{Craig J. Fennie}
 \affiliation{School of Applied and Engineering Physics, Cornell University, Ithaca, New York 14853, USA}
\author{Lena F. Kourkoutis}
  \affiliation{School of Applied and Engineering Physics, Cornell University, Ithaca, New York 14853, USA}
 \affiliation{Kavli Institute at Cornell for Nanoscale Science, Cornell University, Ithaca, New York 14853, USA}

\begin{abstract}
Transition-metal dichalcogenides containing tellurium anions show remarkable charge-lattice modulated structures and prominent interlayer character.
Using cryogenic scanning transmission electron microscopy (STEM), we map the atomic-scale structures of the high temperature (HT) and low temperature (LT) modulated phases in 1T\text{'}-TaTe$_{2}$. 
At HT, we directly show in-plane metal distortions which form trimerized clusters and staggered, three-layer stacking. 
In the LT phase at 93 K, we visualize an additional trimerization of Ta sites and subtle distortions of Te sites by extracting structural information from contrast modulations in plan-view STEM data. 
Coupled with density functional theory calculations and image simulations, this approach opens the door for atomic-scale visualizations of low temperature phase transitions and complex displacements in a variety of layered systems.
\end{abstract}

\maketitle

Many layered transition-metal dichalcogenides (TMD) undergo symmetry-breaking phase transitions by spatially modulating their electronic and structural degrees of freedom.
The modulations may take the form of charge density waves (CDW) and concomitant periodic lattice displacements (PLD) or polymerization of metal sites into one-dimensional (1D) chain structures \cite{Wilson1975}.
These broken-symmetry states couple strongly to electronic behavior, mediating for instance superconductivity or metal-insulator transitions \cite{morosan2006superconductivity,Sipos2008,Kusmartseva2009}. 
Modulated systems also exhibit rich structural complexity whereby the PLD patterns and periodicities vary with the valence of the compound or by external parameters such as temperature and pressure \cite{Wilson1975, Whangbo1992}.
Some well-known superstructures include $\sqrt{13}\times\sqrt{13}$ clustering in 1T-TaS$_{2}$ and 1T-TaSe$_{2}$, 3$\times$3 in 2H-TaSe$_{2}$, and 2$\times$2$\times$2 in 1T-TiSe$_{2}$ \cite{Wilson1975}.

Compared to sulfides or selenides, tellurium-based compounds (MTe$_{2}$, M = V, Nb, Ta, Ir) exhibit distinct electronic and structural features including covalent Te-Te bonds between the layers, enhanced charge transfer to the metal sites and significant in-plane electronic and structural anisotropy \cite{canadell1992importance,Whangbo1992,Jobic1991,Jobic1992,Vernes1998,Chen2017}.
MTe$_{2}$ compounds hosting modulated phases at low temperature thus contain complex interactions between in-plane and out-of-plane ordering tendencies.
In IrTe$_{2}$, for instance, the low temperature phase involves the formation of in-plane Ir-Ir dimers and distortions in interlayer Te bonding \cite{Oh2013, Eom2014, Li2014, Pascut2014}. 

The related 1T'-TaTe$_{2}$ has a distorted, monoclinic structure with space group $C2/m$ at high temperatures \cite{Vernes1998,Sorgel2006,Liu2015}.
The metal atoms form trimerized clusters with short bonds, which are also known as ribbon-chain structures or double zig-zag structures.
Upon cooling, an anomaly appears in the resistivity and magnetic susceptibility at T$_{c}$ $\sim$ 170 K which coincides with the formation of a modulated structure according to X-ray diffraction \cite{Sorgel2006, Chen2017}. 
Pressure and Se substitution were found to destabilize the modulation in favor of superconductivity \cite{Luo2015,Guo2017,Wei2017}. 
In the quest to visualize the local structure, scanning tunneling microscopy (STM) was used to image the high temperature (HT) phase and low temperature (LT) phase on the surface.
Striped contrast in the HT phase was proposed to represent chains of Ta dimers instead of trimers \cite{Feng2016}, however, the stripes are also consistent with the surface Te atoms undergoing the bulk distortion \cite{Kim1997, Chen2018}.
Mapping the bulk structure and disentangling lattice displacements from charge modulations are thus required for resolving the nature of density waves in 1T'-TaTe$_{2}$ and layered TMDs in general \cite{dai2014microscopic,Hildebrand2018, qiao2017anisotropic}. 

In high-angle annular dark-field (HAADF) scanning transmission electron microscopy (STEM), 2D projection images of atomic column positions can be obtained at sub-\AA ngstrom resolution, even under cryogenic conditions \cite{Yankovich2014, ElBaggari2018}. 
Previously, we mapped the superstructure in a charge-ordered manganite at low temperature by directly measuring the picoscale PLD of atomic columns \cite{ElBaggari2018}.
In that particular system, both the atomic structure and the displacements were aligned coherently along the viewing direction.
Layered materials present additional challenges since the superstructures may exhibit periodic and staggered out-of-plane stacking sequences or even disordered stacking.
These interlayer orders are increasingly pertinent for understanding and manipulating the electronic ground states of layered systems \cite{Hovden2016,lee2019origin,Ritschel2018}.
However, the patterns of displacements in the superstructures may be obfuscated by their stacking sequences, requiring atomic-scale probes and new methods that can reveal these distortions.

Here we perform cryogenic HAADF-STEM measurements to visualize the atomic-scale structure of the HT and LT phases in 1T'-TaTe$_{2}$.
We first confirm that the HT phase involves trimer distortions of Ta sites in agreement with earlier refinements of the bulk structure \cite{Vernes1998, Sorgel2006}.
In the LT phase, we observe the formation of a three-fold superstructure, which appears as periodic intensity modulations in STEM data. 
While this LT transition is concealed in cross section \cite{Note1}, analysis of the pattern of the intensity modulations in plan-view achieves a real space visualization of the displacement pattern, consistent with an additional trimerization along the $\mathbf{b}$-axis. 
A combination of density functional theory (DFT) calculations and image simulations confirms the effects of lattice distortions and their staggered stacking on STEM image contrast in plan-view, enabling atomic-scale imaging of transitions with complex lattice order.

\begin{figure}[t]
  \includegraphics[width=.9\linewidth]{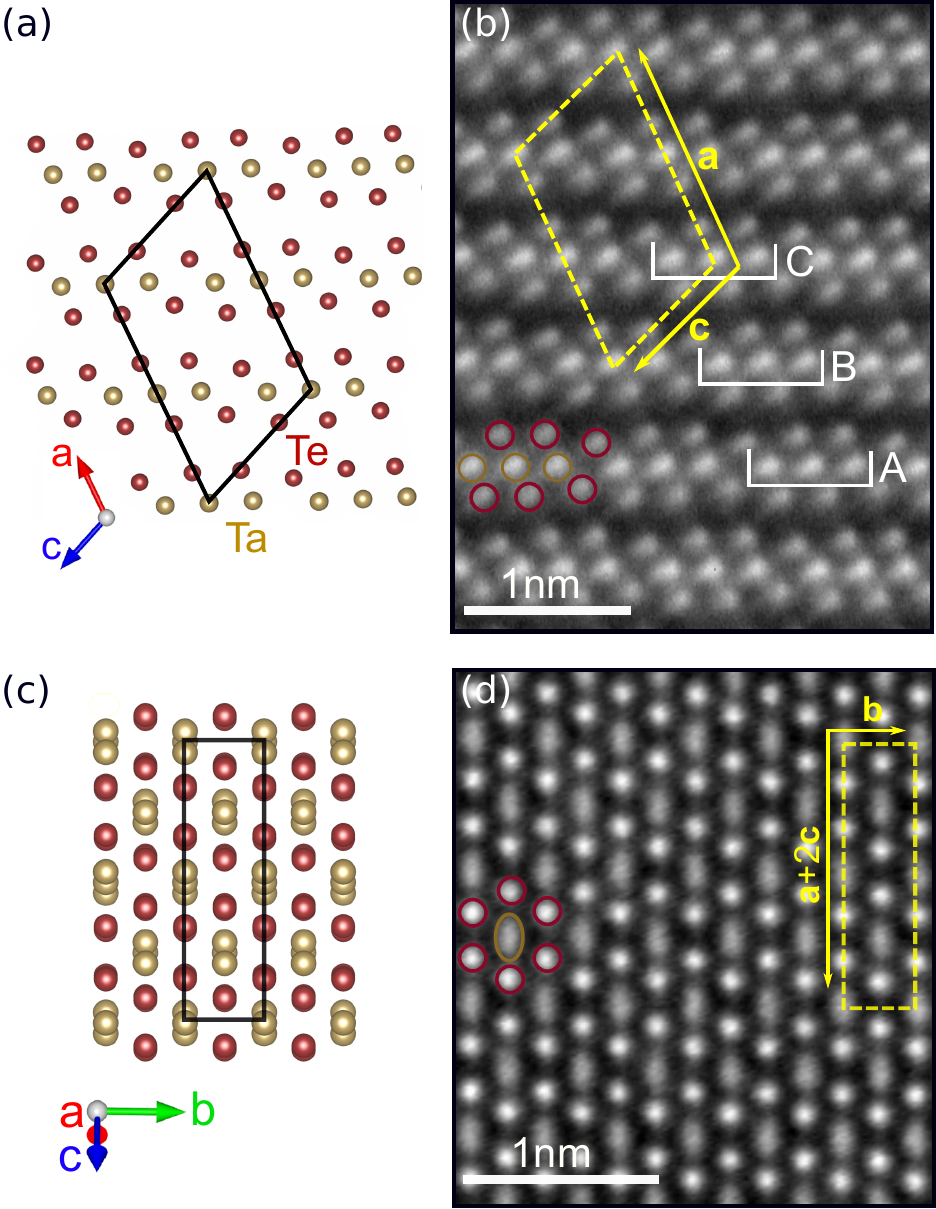}
  \caption{
  (a), (b) Room temperature structural model and HAADF-STEM image of 1T'-TaTe$_{2}$, respectively, viewed in cross section and along the $\mathbf{b}$-direction of the monoclinic cell.
  The unit cell is marked by dashed lines.
  The white brackets delineate trimerized clusters of Ta atoms.
  (c), (d) Structural model and HAADF-STEM image in plan-view at 293 K.
  The Te atoms appear bright and round while Ta atoms appear elongated and dimmer.
  }
  \label{F:Fig1}
\end{figure}

\footnotetext[1]{See Supplemental Material at [URL] for experimental methods, additional cross-sectional data, intensity line profiles, details about the DFT structural optimization, and multislice simulation parameters and results, which includes Refs. \cite{Savitzky2018,Momma2008,Campbell2006,Larsen2017,Allen2015,isotrophy}}
Plan-view samples are prepared by exfoliating commercial bulk crystals (HQ Graphene) onto silicon nitride grids with holes while cross-sectional samples are fabricated using focused ion beam \cite{Note1,stiehl2019current}.
Figures 1(a) and (b) show the room temperature structural model and high-angle annular dark-field (HAADF) STEM image of 1T'-TaTe$_{2}$, viewed along the $\mathbf{b}$-axis of the monoclinic cell.
The contrast in HAADF-STEM scales strongly with the atomic number, $Z$, so Ta columns ($Z$= 73) appear bright and the Te ($Z$ = 52) columns are dimmer.
When the crystal is oriented along a high-symmetry axis, the electron probe can channel along the atomic columns before scattering onto the detector, which further affects the contrast between heavy and light columns \cite{Fitting2006}.  

The room temperature 1T' structure appears highly distorted compared to the 1T phase (space group $P\bar{3}m1$) with one TaTe$_{2}$ formula unit in the unit cell. 
The distortion involves displacements of two Ta columns towards a third stationary Ta column, forming trimerized clusters (white bracket).
The Ta-Ta column distance within a trimer is shortened while the inter-trimer Ta-Ta bond distance is elongated.
By measuring atomic column positions along this viewing direction, we find that the short/long Ta-Ta column distance in projection is 2.76(5)\r{A}/3.97(4)\r{A}, which corresponds to Ta-Ta bond lengths of 3.30(6)\r{A}/4.37(5)\r{A} \cite{Note1}. 
We also observe attendant distortions of tellurium atoms, particularly those that move towards the gap between neighboring trimers.
Moreover, the trimerized clusters are staggered between the layers, forming a three-layer repeat sequence (ABC stacking).
Cross-sectional imaging over larger fields of view does not reveal any regions with AAA stacking or stacking faults, suggesting that the interlayer ordering of the trimers is important \cite{Note1}.  

\footnotetext[2]{The bottom layers in contact with silicon oxide are exposed to air and show defects and amorphous features. The top layer which is exposed under high vaccuum and capped with permalloy maintains the trimerized distortion \cite{Note1,stiehl2019current}.}
In STM visualizations of surface layers in tellurides, distorted chain structures appear as periodic striped contrast reflecting both electronic and lattice contributions.
While the stripes were interpreted as Ta dimer chains by Feng \textit{et al.} \cite{Feng2016}, the contrast more likely reflects the tellurium distortions in the top layer which can occur in both dimerized and trimerized states \cite{Chen2018}.  
Complementing previous refinements of the structure \cite{Kim1997, Sorgel2006}, our data unequivocally shows that 1T'-TaTe$_{2}$ forms trimer distortions. 
Further, these trimers occur throughout the sample including in the top layer of the film \cite{Note2}. 

\footnotetext[3]{Using convergent beam electron diffraction pattern, we estimate the thickness of the specimen to be between 18 and 22 nm (27-33 layers).}
In plan-view, the Ta atoms are misaligned (Fig. 1(c)) due to the ABC stacking of trimers whereas the Te atoms remain stacked on top of each other.
Figure 1(d) shows the STEM image at room temperature in which the Te columns appear round and the Ta columns appear elongated \cite{Note3}.
The elongation, which occurs in the direction perpendicular to the $\mathbf{b}$-axis, reflects the projection nature of STEM.
Furthermore, despite having a larger atomic number, Ta columns are now dimmer than Te columns.
In STEM, the presence of imperfections along the beam direction, such as atomic displacements or chemical disorder, may cause the probe to de-channel leading to measurable changes in the scattered intensity \cite{perovic1993imaging, hillyard1995detector,Fitting2006,Haruta2009,Esser2016}.
The dramatic reduction in Ta column intensity is a direct illustration of how STEM contrast can be sensitive to the 3D arrangement of distorted clusters.

\begin{figure}
  \includegraphics[width=\linewidth]{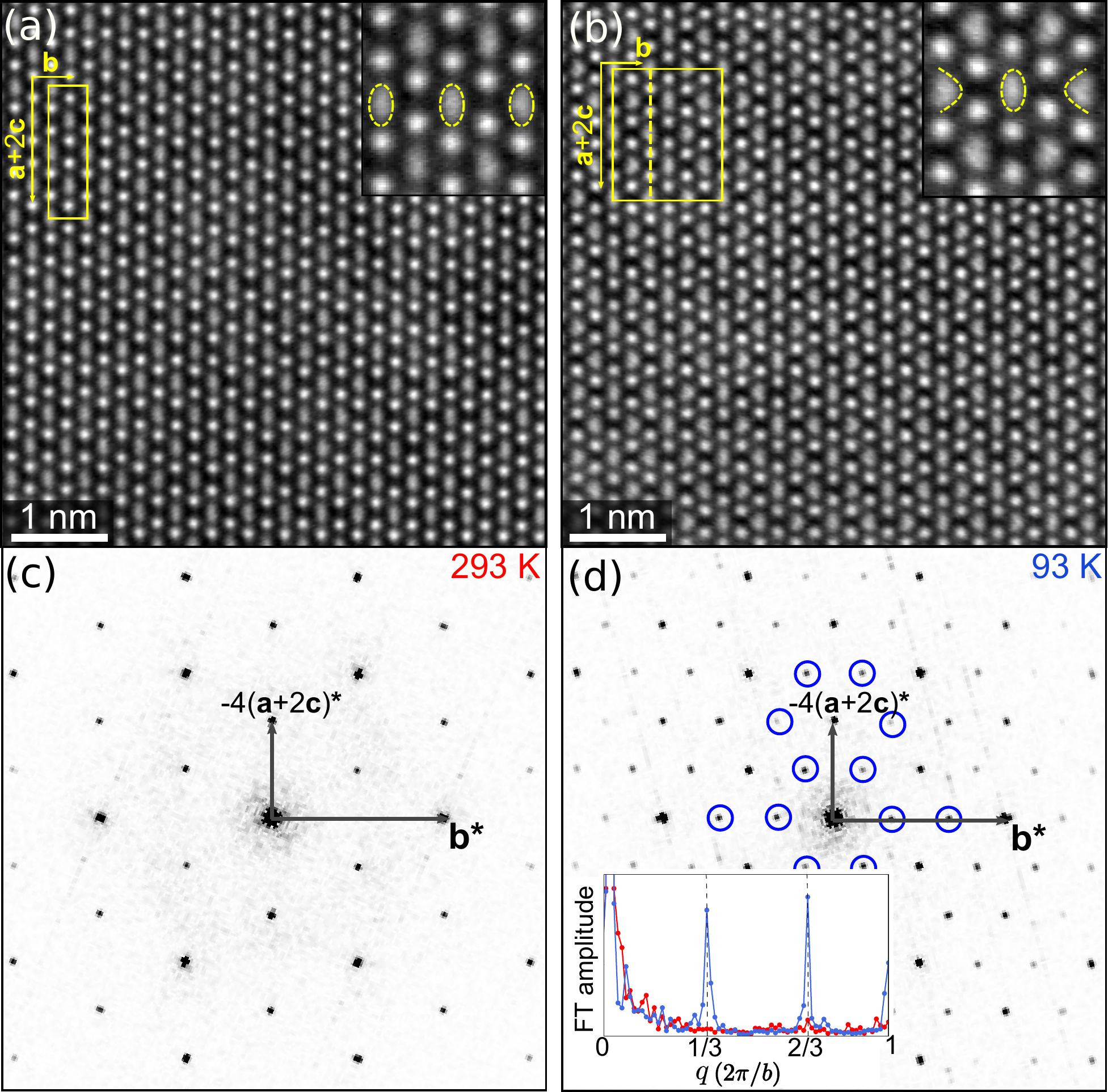}
  \caption{
  (a), (b) STEM image at 293 K and 93 K, respectively.
  The insets show zoomed-in views of the images, over a $\sim$ 1 nm x 1 nm field of view.
  (c), (d) Amplitude of the Fourier transform (FT) at 293 K and 93 K.
  The inset shows linecuts of the FT amplitude from (c) (red) and (d) (blue) along the $\mathbf{b^{*}}$ direction.
  At low temperature, in-plane superlattice peaks (blue circles) with q = 1/3 reciprocal lattice units appear in the FT, reflecting the presence of a modulated structure.  
  }
  \label{F:Fig2}
\end{figure}

Figures 2(a) and (b) show plan-view images at room temperature and at 93 K, respectively.
Before inspecting the real space data, the presence of a modulated structure at low temperature can be confirmed by comparing Fourier transforms (FT) of the two images. 
Figure 2(c) shows the FT amplitude in the HT phase, where the Bragg peaks associated with the crystalline lattice are visible.
The FT amplitude of the LT STEM image (Fig. 2(d)) reveals the presence of superlattice peaks (blue circles).   
The in-plane superlattice wavevector amplitude is q = 1/3 reciprocal lattice units (r.l.u.) (Fig. 2(d), inset), indicating the formation of a three-fold superstructure.

\begin{figure}[t]
  \includegraphics[width=.99\linewidth]{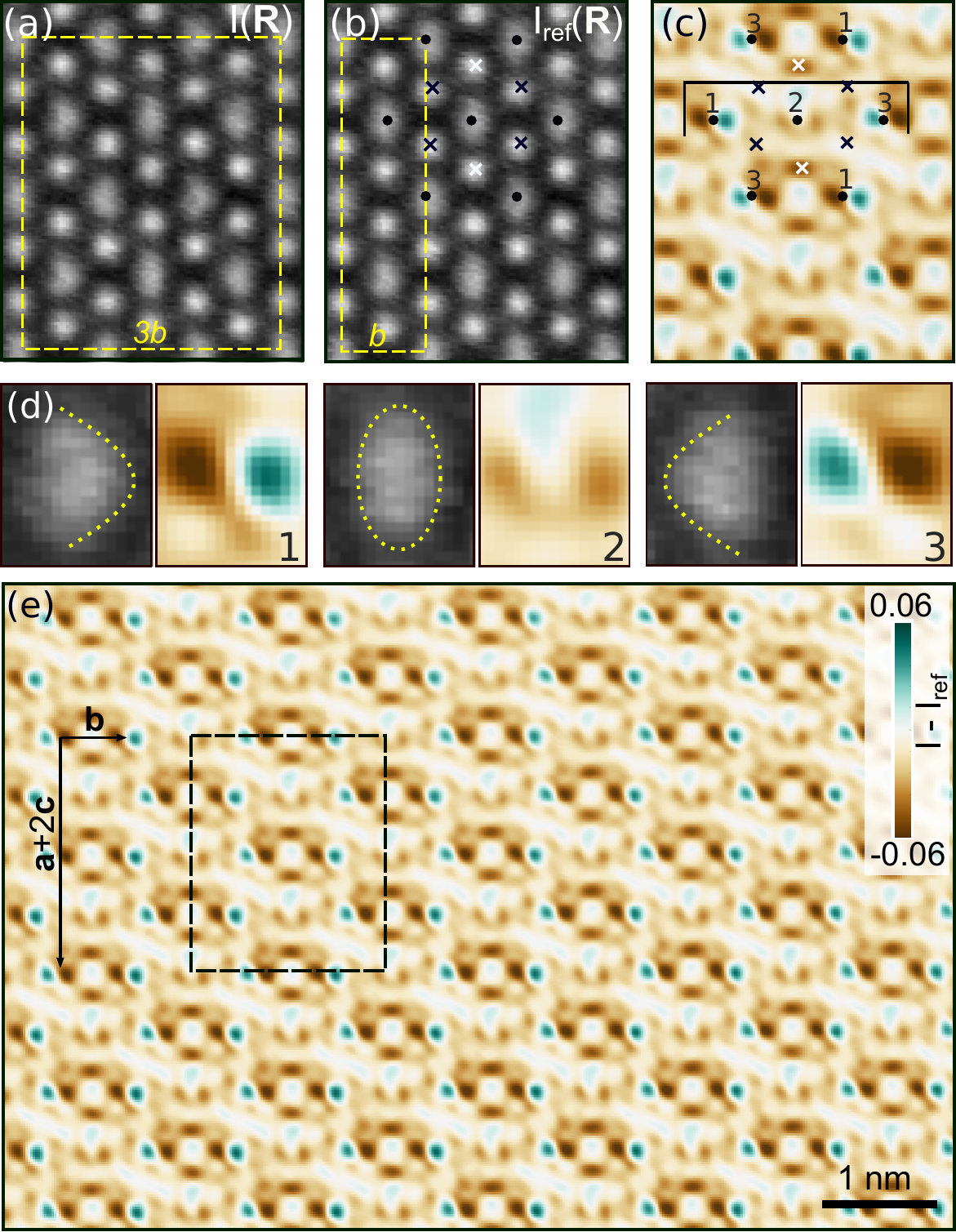}
  \caption{
  (a) Section of the low temperature (LT) STEM image, I(\textbf{R}), spanning a projection of the monoclinic unit cell.
  (b) Generated, reference image, I$_{\text{ref}}$($\textbf{R}$), in the same field of view, as described in the text.
   Black dots represent Ta columns and crosses represent Te columns.
  (c) Difference map between I(\textbf{R}) and I$_{\text{ref}}$($\textbf{R}$).
  The map encodes the modulations in STEM contrast which are responsible for the superlattice peaks.
  The bracket delimits a single low temperature trimerized cluster.
  White crosses represent Te atomic columns that show a reduction in intensity.
  (d) Zoom-in on the Ta-1, Ta-2 and Ta-3 columns and their respective intensity modifications. 
  (e) Large field of view of the difference map showing periodic intensity modulations.
  The LT unit cell is marked by the dashed lines.
  }
  \label{F:Fig3}
\end{figure}

The raw STEM data itself shows structural distortions at low temperature.
The inset in Fig. 2(a) shows a zoomed-in image at room temperature which exhibits the elongation of Ta columns in the HT phase.
Upon cooling, additional deformations are observed in the shapes of Ta columns.
Two out of three Ta columns exhibit moon-shaped distortions, pointing towards a third undistorted Ta column.
By considering the projection nature of STEM, the data indicates that a fraction of Ta atoms in the distorted columns have shifted towards the central Ta site.
In the cross-sectional orientation viewed along the $\mathbf{b}$-axis (as done in Fig. 1(a)), no superlattice is visible at low temperature. 
On the other hand, a low temperature superlattice is detected in the cross-sectional view with the $\mathbf{b}$-axis aligned perpendicular to the beam direction. 
However, this does not reveal the pattern of distortions because the displacements stack in a complicated fashion, necessitating alternative approaches to extract detailed information in plan-view \cite{Note1}.


\begin{figure*}[t]
  \includegraphics[width=.9\textwidth]{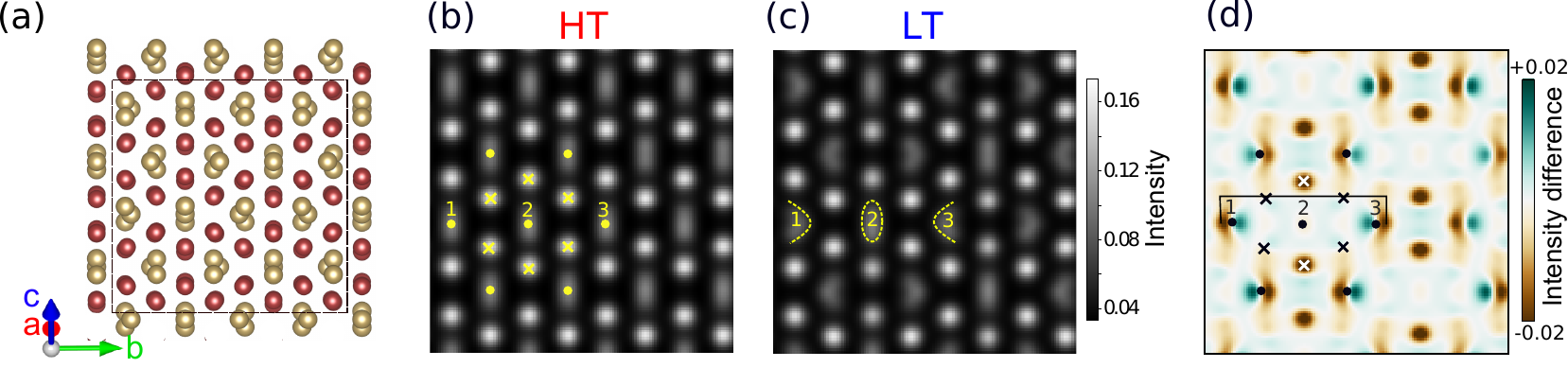}
  \caption{
  (a) Low temperature structure obtained from DFT structural optimization. 
  Tantalum atoms within the existing HT trimers undergo longitudinal periodic lattice displacements along the $\mathbf{b}$-axis, forming additional trimerized clusters.
  More subtle tellurium displacements occur as well.
  The dashed box delimits the simulated field of view.
  (b), (c) Multislice simulation of HAADF-STEM images of the HT and LT phase, respectively.
  The dots represent Ta columns and the crosses represent Te columns.
  The dashed lines highlight distortions in the Ta columns (Ta-1 and Ta-3) at low temperature.
  (d) The difference image shows a periodic intensity modulation that emerges at low temperature.
  The bracket delimits the low temperature Ta trimers and the white crosses represent Te columns that exhibit a reduction in intensity.
  }
  \label{F:Fig4}
\end{figure*}

Ideally, we would like to directly measure the displacements on individual columns, however, the projection nature of STEM requires coherent stacking of both the atoms and their displacements.
For the more complex atomic arrangement in 1T'-TaTe$_{2}$, atom tracking approaches will therefore fail.
To overcome this limitation, we instead analyze HAADF contrast variations which arise from the sensitivity of electron beam channeling to distortions in the lattice.  
The intensity modulations in the image can be extracted by comparing the LT image, I(\textbf{R}), to a reference, un-modulated lattice image, I$_{\text{ref}}$($\textbf{R}$).
We generate such reference image by damping the amplitude of the superlattice peaks in the FT to the background level and applying an inverse FT \cite{savitzky2017bending, ElBaggari2018}. 
The resulting I$_{\text{ref}}$($\textbf{R}$) has no periodic component associated with q = 1/3 r.l.u..
We note that I$_{\text{ref}}$($\textbf{R}$) does not exactly correspond to the HT phase since other distortion modes with different periodicity that may be present remain in the image \cite{Note4}. 

By subtracting the reference lattice image (Fig. 3(b)) from the raw image (Fig. 3(a)), we obtain an intensity map (Fig. 3(c)) which encodes the real space pattern of the intensity modulation at low temperature.
The green (brown) pixels represent an increase (decrease) in HAADF intensity at low temperature. 
The threefold superstructure (black bracket) emerges from  periodic variations in the intensity of the STEM image (Fig. 3(e)).

Focusing on Ta columns (black dots), we observe three kinds of intensity modifications.
The first appears as a dumbbell and involves a decrease of the intensity on the left side of the Ta column and an increase on the right (Fig. 3(d)).
The second, a mirror of the first, involves a decrease of the intensity on the right side of the Ta column and an increase on the left (Fig. 3(d)).
The intensity redistribution of Ta-1 (Ta-3) is consistent with longitudinal displacements of Ta atoms to the right (left) along the $\mathbf{b}$-axis (in the monoclinic cell). 
On Ta-2 columns, the change in intensity is smaller and is consistent throughout the field of view \cite{Note1}.
However, it is not easily interpretable in terms of lattice displacement or stacking disorder.
Together, the low temperature displacement pattern on the Ta columns may be viewed as an additional trimerization along the $\mathbf{b}$-axis \cite{Sorgel2006, Chen2017}.

The superstructure also contains periodic intensity modulations on the Te sites.
In particular, the Te columns which flank Ta-2 columns have lower intensity, which indicates that the stacking of Te atoms is more distorted in the LT phase.
Such tellurium distortions can dramatically affect the electronic behavior of MTe$_{2}$ systems by modulating the interlayer bonding network \cite{canadell1992importance}. 
The Te sites surrounding Ta-1 and Ta-3 columns show negligible changes in STEM intensity. 

We next confirm the real space displacement pattern by comparing to DFT structural optimization.
We use the Vienna \textit{ab-initio} simulation package (VASP) \cite{Kresse1996} code with projector augmented waves \cite{Kresse1999} and the PBEsol exchange-correlation functional \cite{perdew2008restoring} which is known to yield accurate equilibrium structural parameters. 
Starting from the 1T phase of TaTe$_{2}$ (space group $P\bar{3}m1$), we computed phonon frequencies using the supercell frozen-phonon approach \cite{Note1} and found 18 displacement modes with imaginary frequencies, which indicates that the 1T phase is unstable and may lower its energy by condensing combinations of these displacements.
Both the HT and the LT phase were found to have lower energy.
Analysis of the two crystals structures shows that the HT-LT transition involves longitudinal Ta displacements in the Ta-1 and Ta-3 columns which may be described as a trimerization along the $\textbf{b}$-direction, in agreement with the LT distortion measured in the STEM data and in previous structural refinements \cite{Sorgel2006, Chen2017}.
The transition also involves Te distortions as well as zone-center modes.
Remarkably, the difference in energy of the two structures is small despite the large structural distortions in the LT phase.
More details about the DFT methodology, symmetry analysis and atomic positions are given in the supplementary document \cite{Note1}. 

To firmly establish the link between these lattice distortions and STEM contrast modulations at low temperature, we perform multislice simulations on the DFT-optimized structures.
In addition to simulating fully relaxed structures, we simulated images with only Ta or Te displacements, to assign displacement patterns to specific contrast features in the data \cite{Note1}.
Figure 4(b) shows the calculated STEM image in the HT phase, where we observe bright, round tellurium columns and the elongation of tantalum columns, in accordance with the experimental data.
The calculated STEM image based on the fully relaxed LT structure is shown in Fig. 4(c) and the difference between the HT and LT images is shown in Fig. 4(d).
The LT trimerized distortion appears as a deformation of the shape of Ta columns along the $\mathbf{b}$-axis.
In the difference image, the distortion generates the dumbbell-shaped intensity patterns on Ta-1 and Ta-3 columns. 

\footnotetext[4]{Extracting a reference image from the simulated data yields the same dominant intensity modulations associated with the trimer distortion, albeit with some minor differences in the background intensity \cite{Note1}.}
The Ta-2 columns exhibit negligible changes in the contrast whereas the experimental data shows a small change in intensity.
The minor discrepancy might reflect the fact that I$_{\text{ref}}(\mathbf{R})$ does not exactly correspond to the HT phase since other modes present in the structural transition are not captured by the q = 1/3 r.l.u. Fourier components \cite{Note4}.
Alternatively, experimental imperfections such as crystal mis-tilt, aberrations, or surface damage may account for the difference.
That said, the dominant LT contrast modifications on Ta sites are well-explained by a trimerization along the \textbf{b}-direction.
For tellurium columns, both the LT image and the difference image show a reduction in STEM intensity only on the Te sites flanking the Ta-2 column (white crosses), reflecting the variations observed in the experimental data.
Based on the atomic positions in the LT phase, the decrease in STEM intensity on those Te columns is a result of a less coherent stacking of the atoms.
By comparing image simulations and experimental intensity modulations in plan-view, we therefore achieve a visualization of not only the Ta trimerization but also the more subtle Te distortions in the LT phase of 1T'-TaTe$_{2}$.

Probing the bulk structure in layered systems is important for understanding the electronic properties of nominally quasi-2D TMD systems. 
While interlayer interactions are prominent in MTe$_{2}$ compounds, recent studies on 1T-TaS$_{2}$, a prototypical 2D CDW system, also suggest that the out-of-plane ordering of CDW clusters may be responsible for its insulating ground state and its exotic, metastable phases \cite{Ritschel2018}.
These interlayer structures, however, have proven difficult to visualize at atomic resolution \cite{Hildebrand2018,lee2019origin}.
As demonstrated here, mapping contrast modifications in cryogenic STEM enables the visualization of phase transitions with complex displacement patterns.

\begin{acknowledgments}
This work was supported by the National Sciences Foundation (NSF) through the Platform for the Accelerated Realization, Analysis, and Discovery of Interface Materials (DMR-1539918), and made use of the shared facilities of the Cornell Center for Materials Research with funding from the NSF MRSEC program (DMR-1719875). 
Assistance provided by G.M.S. in the preparation of samples was funded by the US Department of Energy (DE-SC0017671).
The FEI Titan Themis 300 was acquired through NSF-MRI-1429155, with additional support from Cornell University, the Weill Institute and the Kavli Institute at Cornell.
\end{acknowledgments}

\bibliography{2019_TaTe2}

\begin{thebibliography}{50}%
\makeatletter
\providecommand \@ifxundefined [1]{%
 \@ifx{#1\undefined}
}%
\providecommand \@ifnum [1]{%
 \ifnum #1\expandafter \@firstoftwo
 \else \expandafter \@secondoftwo
 \fi
}%
\providecommand \@ifx [1]{%
 \ifx #1\expandafter \@firstoftwo
 \else \expandafter \@secondoftwo
 \fi
}%
\providecommand \natexlab [1]{#1}%
\providecommand \enquote  [1]{``#1''}%
\providecommand \bibnamefont  [1]{#1}%
\providecommand \bibfnamefont [1]{#1}%
\providecommand \citenamefont [1]{#1}%
\providecommand \href@noop [0]{\@secondoftwo}%
\providecommand \href [0]{\begingroup \@sanitize@url \@href}%
\providecommand \@href[1]{\@@startlink{#1}\@@href}%
\providecommand \@@href[1]{\endgroup#1\@@endlink}%
\providecommand \@sanitize@url [0]{\catcode `\\12\catcode `\$12\catcode
  `\&12\catcode `\#12\catcode `\^12\catcode `\_12\catcode `\%12\relax}%
\providecommand \@@startlink[1]{}%
\providecommand \@@endlink[0]{}%
\providecommand \url  [0]{\begingroup\@sanitize@url \@url }%
\providecommand \@url [1]{\endgroup\@href {#1}{\urlprefix }}%
\providecommand \urlprefix  [0]{URL }%
\providecommand \Eprint [0]{\href }%
\providecommand \doibase [0]{http://dx.doi.org/}%
\providecommand \selectlanguage [0]{\@gobble}%
\providecommand \bibinfo  [0]{\@secondoftwo}%
\providecommand \bibfield  [0]{\@secondoftwo}%
\providecommand \translation [1]{[#1]}%
\providecommand \BibitemOpen [0]{}%
\providecommand \bibitemStop [0]{}%
\providecommand \bibitemNoStop [0]{.\EOS\space}%
\providecommand \EOS [0]{\spacefactor3000\relax}%
\providecommand \BibitemShut  [1]{\csname bibitem#1\endcsname}%
\let\auto@bib@innerbib\@empty
\bibitem [{\citenamefont {Wilson}\ \emph {et~al.}(1975)\citenamefont {Wilson},
  \citenamefont {{Di Salvo}},\ and\ \citenamefont {Mahajan}}]{Wilson1975}%
  \BibitemOpen
  \bibfield  {author} {\bibinfo {author} {\bibfnamefont {J.}~\bibnamefont
  {Wilson}}, \bibinfo {author} {\bibfnamefont {F.}~\bibnamefont {{Di Salvo}}},
  \ and\ \bibinfo {author} {\bibfnamefont {S.}~\bibnamefont {Mahajan}},\ }\href
  {\doibase 10.1080/00018737500101391} {\bibfield  {journal} {\bibinfo
  {journal} {Advances in Physics}\ }\textbf {\bibinfo {volume} {24}},\ \bibinfo
  {pages} {117} (\bibinfo {year} {1975})}\BibitemShut {NoStop}%
\bibitem [{\citenamefont {Morosan}\ \emph {et~al.}(2006)\citenamefont
  {Morosan}, \citenamefont {Zandbergen}, \citenamefont {Dennis}, \citenamefont
  {Bos}, \citenamefont {Onose}, \citenamefont {Klimczuk}, \citenamefont
  {Ramirez}, \citenamefont {Ong},\ and\ \citenamefont
  {Cava}}]{morosan2006superconductivity}%
  \BibitemOpen
  \bibfield  {author} {\bibinfo {author} {\bibfnamefont {E.}~\bibnamefont
  {Morosan}}, \bibinfo {author} {\bibfnamefont {H.~W.}\ \bibnamefont
  {Zandbergen}}, \bibinfo {author} {\bibfnamefont {B.~S.}\ \bibnamefont
  {Dennis}}, \bibinfo {author} {\bibfnamefont {J.~W.~G.}\ \bibnamefont {Bos}},
  \bibinfo {author} {\bibfnamefont {Y.}~\bibnamefont {Onose}}, \bibinfo
  {author} {\bibfnamefont {T.}~\bibnamefont {Klimczuk}}, \bibinfo {author}
  {\bibfnamefont {A.~P.}\ \bibnamefont {Ramirez}}, \bibinfo {author}
  {\bibfnamefont {N.~P.}\ \bibnamefont {Ong}}, \ and\ \bibinfo {author}
  {\bibfnamefont {R.~J.}\ \bibnamefont {Cava}},\ }\href@noop {} {\bibfield
  {journal} {\bibinfo  {journal} {Nature Physics}\ }\textbf {\bibinfo {volume}
  {2}},\ \bibinfo {pages} {544} (\bibinfo {year} {2006})}\BibitemShut {NoStop}%
\bibitem [{\citenamefont {Sipos}\ \emph {et~al.}(2008)\citenamefont {Sipos},
  \citenamefont {Kusmartseva}, \citenamefont {Akrap}, \citenamefont {Berger},
  \citenamefont {Forr{\'{o}}},\ and\ \citenamefont {Tuti\v{s}}}]{Sipos2008}%
  \BibitemOpen
  \bibfield  {author} {\bibinfo {author} {\bibfnamefont {B.}~\bibnamefont
  {Sipos}}, \bibinfo {author} {\bibfnamefont {A.~F.}\ \bibnamefont
  {Kusmartseva}}, \bibinfo {author} {\bibfnamefont {A.}~\bibnamefont {Akrap}},
  \bibinfo {author} {\bibfnamefont {H.}~\bibnamefont {Berger}}, \bibinfo
  {author} {\bibfnamefont {L.}~\bibnamefont {Forr{\'{o}}}}, \ and\ \bibinfo
  {author} {\bibfnamefont {E.}~\bibnamefont {Tuti\v{s}}},\ }\href {\doibase
  10.1038/nmat2318} {\bibfield  {journal} {\bibinfo  {journal} {Nature
  materials}\ }\textbf {\bibinfo {volume} {7}},\ \bibinfo {pages} {960}
  (\bibinfo {year} {2008})}\BibitemShut {NoStop}%
\bibitem [{\citenamefont {Kusmartseva}\ \emph {et~al.}(2009)\citenamefont
  {Kusmartseva}, \citenamefont {Sipos}, \citenamefont {Berger}, \citenamefont
  {Forr\'{o}},\ and\ \citenamefont {Tuti\v{s}}}]{Kusmartseva2009}%
  \BibitemOpen
  \bibfield  {author} {\bibinfo {author} {\bibfnamefont {A.~F.}\ \bibnamefont
  {Kusmartseva}}, \bibinfo {author} {\bibfnamefont {B.}~\bibnamefont {Sipos}},
  \bibinfo {author} {\bibfnamefont {H.}~\bibnamefont {Berger}}, \bibinfo
  {author} {\bibfnamefont {L.}~\bibnamefont {Forr\'{o}}}, \ and\ \bibinfo
  {author} {\bibfnamefont {E.}~\bibnamefont {Tuti\v{s}}},\ }\href {\doibase
  10.1103/PhysRevLett.103.236401} {\bibfield  {journal} {\bibinfo  {journal}
  {Physical Review Letters}\ }\textbf {\bibinfo {volume} {103}},\ \bibinfo
  {pages} {236401} (\bibinfo {year} {2009})}\BibitemShut {NoStop}%
\bibitem [{\citenamefont {Whangbo}\ and\ \citenamefont
  {Canadell}(1992)}]{Whangbo1992}%
  \BibitemOpen
  \bibfield  {author} {\bibinfo {author} {\bibfnamefont {M.~H.}\ \bibnamefont
  {Whangbo}}\ and\ \bibinfo {author} {\bibfnamefont {E.}~\bibnamefont
  {Canadell}},\ }\href {\doibase 10.1021/ja00050a044} {\bibfield  {journal}
  {\bibinfo  {journal} {Journal of the American Chemical Society}\ }\textbf
  {\bibinfo {volume} {114}},\ \bibinfo {pages} {9587} (\bibinfo {year}
  {1992})}\BibitemShut {NoStop}%
\bibitem [{\citenamefont {Canadell}\ \emph {et~al.}(1992)\citenamefont
  {Canadell}, \citenamefont {Brec}, \citenamefont {Rouxel}, \citenamefont
  {Whangbo} \emph {et~al.}}]{canadell1992importance}%
  \BibitemOpen
  \bibfield  {author} {\bibinfo {author} {\bibfnamefont {E.}~\bibnamefont
  {Canadell}}, \bibinfo {author} {\bibfnamefont {R.}~\bibnamefont {Brec}},
  \bibinfo {author} {\bibfnamefont {J.}~\bibnamefont {Rouxel}}, \bibinfo
  {author} {\bibfnamefont {M.-H.}\ \bibnamefont {Whangbo}},  \emph {et~al.},\
  }\href@noop {} {\bibfield  {journal} {\bibinfo  {journal} {Journal of Solid
  State Chemistry}\ }\textbf {\bibinfo {volume} {99}},\ \bibinfo {pages} {189}
  (\bibinfo {year} {1992})}\BibitemShut {NoStop}%
\bibitem [{\citenamefont {Jobic}\ \emph {et~al.}(1991)\citenamefont {Jobic},
  \citenamefont {Deniard}, \citenamefont {Brec}, \citenamefont {Rouxel},
  \citenamefont {Jouanneaux},\ and\ \citenamefont {Fitch}}]{Jobic1991}%
  \BibitemOpen
  \bibfield  {author} {\bibinfo {author} {\bibfnamefont {S.}~\bibnamefont
  {Jobic}}, \bibinfo {author} {\bibfnamefont {P.}~\bibnamefont {Deniard}},
  \bibinfo {author} {\bibfnamefont {R.}~\bibnamefont {Brec}}, \bibinfo {author}
  {\bibfnamefont {J.}~\bibnamefont {Rouxel}}, \bibinfo {author} {\bibfnamefont
  {A.}~\bibnamefont {Jouanneaux}}, \ and\ \bibinfo {author} {\bibfnamefont
  {A.~N.}\ \bibnamefont {Fitch}},\ }\href {\doibase 10.1002/zaac.19915980119}
  {\bibfield  {journal} {\bibinfo  {journal} {ZAAC ‐ Journal of Inorganic and
  General Chemistry}\ }\textbf {\bibinfo {volume} {598}},\ \bibinfo {pages}
  {199} (\bibinfo {year} {1991})}\BibitemShut {NoStop}%
\bibitem [{\citenamefont {Jobic}\ \emph {et~al.}(1992)\citenamefont {Jobic},
  \citenamefont {Brec},\ and\ \citenamefont {Rouxel}}]{Jobic1992}%
  \BibitemOpen
  \bibfield  {author} {\bibinfo {author} {\bibfnamefont {S.}~\bibnamefont
  {Jobic}}, \bibinfo {author} {\bibfnamefont {R.}~\bibnamefont {Brec}}, \ and\
  \bibinfo {author} {\bibfnamefont {J.}~\bibnamefont {Rouxel}},\ }\href
  {\doibase 10.1016/S0022-4596(05)80309-3} {\bibfield  {journal} {\bibinfo
  {journal} {Journal of Solid State Chemistry}\ }\textbf {\bibinfo {volume}
  {96}},\ \bibinfo {pages} {169} (\bibinfo {year} {1992})}\BibitemShut
  {NoStop}%
\bibitem [{\citenamefont {Vernes}\ \emph {et~al.}(1998)\citenamefont {Vernes},
  \citenamefont {Ebert}, \citenamefont {Bensch}, \citenamefont {Heid},\ and\
  \citenamefont {N{\"{a}}ther}}]{Vernes1998}%
  \BibitemOpen
  \bibfield  {author} {\bibinfo {author} {\bibfnamefont {A.}~\bibnamefont
  {Vernes}}, \bibinfo {author} {\bibfnamefont {H.}~\bibnamefont {Ebert}},
  \bibinfo {author} {\bibfnamefont {W.}~\bibnamefont {Bensch}}, \bibinfo
  {author} {\bibfnamefont {W.}~\bibnamefont {Heid}}, \ and\ \bibinfo {author}
  {\bibfnamefont {C.}~\bibnamefont {N{\"{a}}ther}},\ }\href {\doibase
  10.1088/0953-8984/10/4/006} {\bibfield  {journal} {\bibinfo  {journal}
  {Journal of Physics Condensed Matter}\ }\textbf {\bibinfo {volume} {10}},\
  \bibinfo {pages} {761} (\bibinfo {year} {1998})}\BibitemShut {NoStop}%
\bibitem [{\citenamefont {Chen}\ \emph {et~al.}(2017)\citenamefont {Chen},
  \citenamefont {Li}, \citenamefont {Fan}, \citenamefont {Guo},\ and\
  \citenamefont {Chen}}]{Chen2017}%
  \BibitemOpen
  \bibfield  {author} {\bibinfo {author} {\bibfnamefont {H.}~\bibnamefont
  {Chen}}, \bibinfo {author} {\bibfnamefont {Z.}~\bibnamefont {Li}}, \bibinfo
  {author} {\bibfnamefont {X.}~\bibnamefont {Fan}}, \bibinfo {author}
  {\bibfnamefont {L.}~\bibnamefont {Guo}}, \ and\ \bibinfo {author}
  {\bibfnamefont {X.}~\bibnamefont {Chen}},\ }\href
  {http://arxiv.org/abs/1706.08661} {\ \textbf {\bibinfo {volume} {2}}
  (\bibinfo {year} {2017})},\ \Eprint {http://arxiv.org/abs/1706.08661}
  {arXiv:1706.08661} \BibitemShut {NoStop}%
\bibitem [{\citenamefont {Oh}\ \emph {et~al.}(2013)\citenamefont {Oh},
  \citenamefont {Yang}, \citenamefont {Horibe},\ and\ \citenamefont
  {Cheong}}]{Oh2013}%
  \BibitemOpen
  \bibfield  {author} {\bibinfo {author} {\bibfnamefont {Y.~S.}\ \bibnamefont
  {Oh}}, \bibinfo {author} {\bibfnamefont {J.~J.}\ \bibnamefont {Yang}},
  \bibinfo {author} {\bibfnamefont {Y.}~\bibnamefont {Horibe}}, \ and\ \bibinfo
  {author} {\bibfnamefont {S.-W.}\ \bibnamefont {Cheong}},\ }\href@noop {}
  {\bibfield  {journal} {\bibinfo  {journal} {Physical review letters}\
  }\textbf {\bibinfo {volume} {110}},\ \bibinfo {pages} {127209} (\bibinfo
  {year} {2013})}\BibitemShut {NoStop}%
\bibitem [{\citenamefont {Eom}\ \emph {et~al.}(2014)\citenamefont {Eom},
  \citenamefont {Kim}, \citenamefont {Jo}, \citenamefont {Yang}, \citenamefont
  {Choi}, \citenamefont {Min}, \citenamefont {Park}, \citenamefont {Cheong},\
  and\ \citenamefont {Kim}}]{Eom2014}%
  \BibitemOpen
  \bibfield  {author} {\bibinfo {author} {\bibfnamefont {M.~J.}\ \bibnamefont
  {Eom}}, \bibinfo {author} {\bibfnamefont {K.}~\bibnamefont {Kim}}, \bibinfo
  {author} {\bibfnamefont {Y.~J.}\ \bibnamefont {Jo}}, \bibinfo {author}
  {\bibfnamefont {J.~J.}\ \bibnamefont {Yang}}, \bibinfo {author}
  {\bibfnamefont {E.~S.}\ \bibnamefont {Choi}}, \bibinfo {author}
  {\bibfnamefont {B.~I.}\ \bibnamefont {Min}}, \bibinfo {author} {\bibfnamefont
  {J.-H.}\ \bibnamefont {Park}}, \bibinfo {author} {\bibfnamefont {S.-W.}\
  \bibnamefont {Cheong}}, \ and\ \bibinfo {author} {\bibfnamefont {J.~S.}\
  \bibnamefont {Kim}},\ }\href@noop {} {\bibfield  {journal} {\bibinfo
  {journal} {Physical review letters}\ }\textbf {\bibinfo {volume} {113}},\
  \bibinfo {pages} {266406} (\bibinfo {year} {2014})}\BibitemShut {NoStop}%
\bibitem [{\citenamefont {Li}\ \emph {et~al.}(2014)\citenamefont {Li},
  \citenamefont {Lin}, \citenamefont {Yan}, \citenamefont {Chen}, \citenamefont
  {Gianfrancesco}, \citenamefont {Singh}, \citenamefont {Mandrus},
  \citenamefont {Kalinin},\ and\ \citenamefont {Pan}}]{Li2014}%
  \BibitemOpen
  \bibfield  {author} {\bibinfo {author} {\bibfnamefont {Q.}~\bibnamefont
  {Li}}, \bibinfo {author} {\bibfnamefont {W.}~\bibnamefont {Lin}}, \bibinfo
  {author} {\bibfnamefont {J.}~\bibnamefont {Yan}}, \bibinfo {author}
  {\bibfnamefont {X.}~\bibnamefont {Chen}}, \bibinfo {author} {\bibfnamefont
  {A.~G.}\ \bibnamefont {Gianfrancesco}}, \bibinfo {author} {\bibfnamefont
  {D.~J.}\ \bibnamefont {Singh}}, \bibinfo {author} {\bibfnamefont
  {D.}~\bibnamefont {Mandrus}}, \bibinfo {author} {\bibfnamefont {S.~V.}\
  \bibnamefont {Kalinin}}, \ and\ \bibinfo {author} {\bibfnamefont
  {M.}~\bibnamefont {Pan}},\ }\href {\doibase 10.1038/ncomms6358} {\bibfield
  {journal} {\bibinfo  {journal} {Nature Communications}\ }\textbf {\bibinfo
  {volume} {5}},\ \bibinfo {pages} {1} (\bibinfo {year} {2014})}\BibitemShut
  {NoStop}%
\bibitem [{\citenamefont {Pascut}\ \emph {et~al.}(2014)\citenamefont {Pascut},
  \citenamefont {Birol}, \citenamefont {Gutmann}, \citenamefont {Yang},
  \citenamefont {Cheong}, \citenamefont {Haule},\ and\ \citenamefont
  {Kiryukhin}}]{Pascut2014}%
  \BibitemOpen
  \bibfield  {author} {\bibinfo {author} {\bibfnamefont {G.~L.}\ \bibnamefont
  {Pascut}}, \bibinfo {author} {\bibfnamefont {T.}~\bibnamefont {Birol}},
  \bibinfo {author} {\bibfnamefont {M.~J.}\ \bibnamefont {Gutmann}}, \bibinfo
  {author} {\bibfnamefont {J.~J.}\ \bibnamefont {Yang}}, \bibinfo {author}
  {\bibfnamefont {S.-W.}\ \bibnamefont {Cheong}}, \bibinfo {author}
  {\bibfnamefont {K.}~\bibnamefont {Haule}}, \ and\ \bibinfo {author}
  {\bibfnamefont {V.}~\bibnamefont {Kiryukhin}},\ }\href@noop {} {\bibfield
  {journal} {\bibinfo  {journal} {Physical Review B}\ }\textbf {\bibinfo
  {volume} {90}},\ \bibinfo {pages} {195122} (\bibinfo {year}
  {2014})}\BibitemShut {NoStop}%
\bibitem [{\citenamefont {S\"{o}rgel}\ \emph {et~al.}(2006)\citenamefont
  {S\"{o}rgel}, \citenamefont {Nuss}, \citenamefont {Wedig}, \citenamefont
  {Kremer},\ and\ \citenamefont {Jansen}}]{Sorgel2006}%
  \BibitemOpen
  \bibfield  {author} {\bibinfo {author} {\bibfnamefont {T.}~\bibnamefont
  {S\"{o}rgel}}, \bibinfo {author} {\bibfnamefont {J.}~\bibnamefont {Nuss}},
  \bibinfo {author} {\bibfnamefont {U.}~\bibnamefont {Wedig}}, \bibinfo
  {author} {\bibfnamefont {R.}~\bibnamefont {Kremer}}, \ and\ \bibinfo {author}
  {\bibfnamefont {M.}~\bibnamefont {Jansen}},\ }\href {\doibase
  10.1016/j.materresbull.2006.02.020} {\bibfield  {journal} {\bibinfo
  {journal} {Materials Research Bulletin}\ }\textbf {\bibinfo {volume} {41}},\
  \bibinfo {pages} {987} (\bibinfo {year} {2006})}\BibitemShut {NoStop}%
\bibitem [{\citenamefont {Liu}\ \emph {et~al.}(2015)\citenamefont {Liu},
  \citenamefont {Lu}, \citenamefont {Shao}, \citenamefont {Zu}, \citenamefont
  {Kan}, \citenamefont {Song},\ and\ \citenamefont {Sun}}]{Liu2015}%
  \BibitemOpen
  \bibfield  {author} {\bibinfo {author} {\bibfnamefont {Y.}~\bibnamefont
  {Liu}}, \bibinfo {author} {\bibfnamefont {W.~J.}\ \bibnamefont {Lu}},
  \bibinfo {author} {\bibfnamefont {D.~F.}\ \bibnamefont {Shao}}, \bibinfo
  {author} {\bibfnamefont {L.}~\bibnamefont {Zu}}, \bibinfo {author}
  {\bibfnamefont {X.~C.}\ \bibnamefont {Kan}}, \bibinfo {author} {\bibfnamefont
  {W.~H.}\ \bibnamefont {Song}}, \ and\ \bibinfo {author} {\bibfnamefont
  {Y.~P.}\ \bibnamefont {Sun}},\ }\href {\doibase 10.1209/0295-5075/109/17003}
  {\bibfield  {journal} {\bibinfo  {journal} {EPL (Europhysics Letters)}\
  }\textbf {\bibinfo {volume} {109}},\ \bibinfo {pages} {17003} (\bibinfo
  {year} {2015})}\BibitemShut {NoStop}%
\bibitem [{\citenamefont {Luo}\ \emph {et~al.}(2015)\citenamefont {Luo},
  \citenamefont {Xie}, \citenamefont {Tao}, \citenamefont {Inoue},
  \citenamefont {Gyenis}, \citenamefont {Krizan}, \citenamefont {Yazdani},
  \citenamefont {Zhu},\ and\ \citenamefont {Cava}}]{Luo2015}%
  \BibitemOpen
  \bibfield  {author} {\bibinfo {author} {\bibfnamefont {H.}~\bibnamefont
  {Luo}}, \bibinfo {author} {\bibfnamefont {W.}~\bibnamefont {Xie}}, \bibinfo
  {author} {\bibfnamefont {J.}~\bibnamefont {Tao}}, \bibinfo {author}
  {\bibfnamefont {H.}~\bibnamefont {Inoue}}, \bibinfo {author} {\bibfnamefont
  {A.}~\bibnamefont {Gyenis}}, \bibinfo {author} {\bibfnamefont {J.~W.}\
  \bibnamefont {Krizan}}, \bibinfo {author} {\bibfnamefont {A.}~\bibnamefont
  {Yazdani}}, \bibinfo {author} {\bibfnamefont {Y.}~\bibnamefont {Zhu}}, \ and\
  \bibinfo {author} {\bibfnamefont {R.~J.}\ \bibnamefont {Cava}},\ }\href
  {\doibase 10.1073/pnas.1502460112} {\bibfield  {journal} {\bibinfo  {journal}
  {Proceedings of the National Academy of Sciences}\ }\textbf {\bibinfo
  {volume} {112}},\ \bibinfo {pages} {E1174} (\bibinfo {year}
  {2015})}\BibitemShut {NoStop}%
\bibitem [{\citenamefont {Guo}\ \emph {et~al.}(2017)\citenamefont {Guo},
  \citenamefont {Luo}, \citenamefont {Yang}, \citenamefont {Wei}, \citenamefont
  {Wang}, \citenamefont {Yi}, \citenamefont {Zhou}, \citenamefont {Wang},
  \citenamefont {Cai}, \citenamefont {Zhang}, \citenamefont {Li}, \citenamefont
  {Li}, \citenamefont {Liu}, \citenamefont {Yang}, \citenamefont {Li},
  \citenamefont {Li}, \citenamefont {Wu}, \citenamefont {Cava},\ and\
  \citenamefont {Sun}}]{Guo2017}%
  \BibitemOpen
  \bibfield  {author} {\bibinfo {author} {\bibfnamefont {J.}~\bibnamefont
  {Guo}}, \bibinfo {author} {\bibfnamefont {H.}~\bibnamefont {Luo}}, \bibinfo
  {author} {\bibfnamefont {H.}~\bibnamefont {Yang}}, \bibinfo {author}
  {\bibfnamefont {L.}~\bibnamefont {Wei}}, \bibinfo {author} {\bibfnamefont
  {H.}~\bibnamefont {Wang}}, \bibinfo {author} {\bibfnamefont {W.}~\bibnamefont
  {Yi}}, \bibinfo {author} {\bibfnamefont {Y.}~\bibnamefont {Zhou}}, \bibinfo
  {author} {\bibfnamefont {Z.}~\bibnamefont {Wang}}, \bibinfo {author}
  {\bibfnamefont {S.}~\bibnamefont {Cai}}, \bibinfo {author} {\bibfnamefont
  {S.}~\bibnamefont {Zhang}}, \bibinfo {author} {\bibfnamefont
  {X.}~\bibnamefont {Li}}, \bibinfo {author} {\bibfnamefont {Y.}~\bibnamefont
  {Li}}, \bibinfo {author} {\bibfnamefont {J.}~\bibnamefont {Liu}}, \bibinfo
  {author} {\bibfnamefont {K.}~\bibnamefont {Yang}}, \bibinfo {author}
  {\bibfnamefont {A.}~\bibnamefont {Li}}, \bibinfo {author} {\bibfnamefont
  {J.}~\bibnamefont {Li}}, \bibinfo {author} {\bibfnamefont {Q.}~\bibnamefont
  {Wu}}, \bibinfo {author} {\bibfnamefont {R.~J.}\ \bibnamefont {Cava}}, \ and\
  \bibinfo {author} {\bibfnamefont {L.}~\bibnamefont {Sun}},\ }\href
  {https://arxiv.org/pdf/1704.08106.pdf http://arxiv.org/abs/1704.08106} {\
  (\bibinfo {year} {2017})},\ \Eprint {http://arxiv.org/abs/1704.08106}
  {arXiv:1704.08106} \BibitemShut {NoStop}%
\bibitem [{\citenamefont {Wei}\ \emph {et~al.}(2017)\citenamefont {Wei},
  \citenamefont {Sun}, \citenamefont {Sun}, \citenamefont {Liu}, \citenamefont
  {Shao}, \citenamefont {Lu}, \citenamefont {Sun}, \citenamefont {Tian},\ and\
  \citenamefont {Yang}}]{Wei2017}%
  \BibitemOpen
  \bibfield  {author} {\bibinfo {author} {\bibfnamefont {L.~L.}\ \bibnamefont
  {Wei}}, \bibinfo {author} {\bibfnamefont {S.~S.}\ \bibnamefont {Sun}},
  \bibinfo {author} {\bibfnamefont {K.}~\bibnamefont {Sun}}, \bibinfo {author}
  {\bibfnamefont {Y.}~\bibnamefont {Liu}}, \bibinfo {author} {\bibfnamefont
  {D.~F.}\ \bibnamefont {Shao}}, \bibinfo {author} {\bibfnamefont {W.~J.}\
  \bibnamefont {Lu}}, \bibinfo {author} {\bibfnamefont {Y.~P.}\ \bibnamefont
  {Sun}}, \bibinfo {author} {\bibfnamefont {H.~F.}\ \bibnamefont {Tian}}, \
  and\ \bibinfo {author} {\bibfnamefont {H.~X.}\ \bibnamefont {Yang}},\ }\href
  {\doibase 10.1088/0256-307X/34/8/086101} {\bibfield  {journal} {\bibinfo
  {journal} {Chinese Physics Letters}\ }\textbf {\bibinfo {volume} {34}}
  (\bibinfo {year} {2017}),\ 10.1088/0256-307X/34/8/086101}\BibitemShut
  {NoStop}%
\bibitem [{\citenamefont {Feng}\ \emph {et~al.}(2016)\citenamefont {Feng},
  \citenamefont {Tan}, \citenamefont {Wagner}, \citenamefont {Liu},
  \citenamefont {Mao}, \citenamefont {Ke},\ and\ \citenamefont
  {Zhang}}]{Feng2016}%
  \BibitemOpen
  \bibfield  {author} {\bibinfo {author} {\bibfnamefont {J.}~\bibnamefont
  {Feng}}, \bibinfo {author} {\bibfnamefont {A.}~\bibnamefont {Tan}}, \bibinfo
  {author} {\bibfnamefont {S.}~\bibnamefont {Wagner}}, \bibinfo {author}
  {\bibfnamefont {J.}~\bibnamefont {Liu}}, \bibinfo {author} {\bibfnamefont
  {Z.}~\bibnamefont {Mao}}, \bibinfo {author} {\bibfnamefont {X.}~\bibnamefont
  {Ke}}, \ and\ \bibinfo {author} {\bibfnamefont {P.}~\bibnamefont {Zhang}},\
  }\href {\doibase 10.1063/1.4958616} {\bibfield  {journal} {\bibinfo
  {journal} {Applied Physics Letters}\ }\textbf {\bibinfo {volume} {109}}
  (\bibinfo {year} {2016}),\ 10.1063/1.4958616}\BibitemShut {NoStop}%
\bibitem [{\citenamefont {Kim}\ \emph {et~al.}(1997)\citenamefont {Kim},
  \citenamefont {Park}, \citenamefont {Jeon}, \citenamefont {Kim},
  \citenamefont {Pyun},\ and\ \citenamefont {Yee}}]{Kim1997}%
  \BibitemOpen
  \bibfield  {author} {\bibinfo {author} {\bibfnamefont {S.-J.}\ \bibnamefont
  {Kim}}, \bibinfo {author} {\bibfnamefont {S.-J.}\ \bibnamefont {Park}},
  \bibinfo {author} {\bibfnamefont {I.~C.}\ \bibnamefont {Jeon}}, \bibinfo
  {author} {\bibfnamefont {C.}~\bibnamefont {Kim}}, \bibinfo {author}
  {\bibfnamefont {C.}~\bibnamefont {Pyun}}, \ and\ \bibinfo {author}
  {\bibfnamefont {K.~A.}\ \bibnamefont {Yee}},\ }\href {\doibase
  10.1016/S0022-3697(96)00146-1} {\bibfield  {journal} {\bibinfo  {journal}
  {Journal of Physics and Chemistry of Solids}\ }\textbf {\bibinfo {volume}
  {58}},\ \bibinfo {pages} {659} (\bibinfo {year} {1997})}\BibitemShut
  {NoStop}%
\bibitem [{\citenamefont {Chen}\ \emph {et~al.}(2018)\citenamefont {Chen},
  \citenamefont {Kim}, \citenamefont {Admasu}, \citenamefont {Cheong},
  \citenamefont {Haule}, \citenamefont {Vanderbilt},\ and\ \citenamefont
  {Wu}}]{Chen2018}%
  \BibitemOpen
  \bibfield  {author} {\bibinfo {author} {\bibfnamefont {C.}~\bibnamefont
  {Chen}}, \bibinfo {author} {\bibfnamefont {H.-S.}\ \bibnamefont {Kim}},
  \bibinfo {author} {\bibfnamefont {A.~S.}\ \bibnamefont {Admasu}}, \bibinfo
  {author} {\bibfnamefont {S.-W.}\ \bibnamefont {Cheong}}, \bibinfo {author}
  {\bibfnamefont {K.}~\bibnamefont {Haule}}, \bibinfo {author} {\bibfnamefont
  {D.}~\bibnamefont {Vanderbilt}}, \ and\ \bibinfo {author} {\bibfnamefont
  {W.}~\bibnamefont {Wu}},\ }\href {\doibase 10.1103/PhysRevB.98.195423}
  {\bibfield  {journal} {\bibinfo  {journal} {Physical Review B}\ }\textbf
  {\bibinfo {volume} {98}},\ \bibinfo {pages} {195423} (\bibinfo {year}
  {2018})}\BibitemShut {NoStop}%
\bibitem [{\citenamefont {Dai}\ \emph {et~al.}(2014)\citenamefont {Dai},
  \citenamefont {Calleja}, \citenamefont {Alldredge}, \citenamefont {Zhu},
  \citenamefont {Li}, \citenamefont {Lu}, \citenamefont {Sun}, \citenamefont
  {Wolf}, \citenamefont {Berger},\ and\ \citenamefont
  {McElroy}}]{dai2014microscopic}%
  \BibitemOpen
  \bibfield  {author} {\bibinfo {author} {\bibfnamefont {J.}~\bibnamefont
  {Dai}}, \bibinfo {author} {\bibfnamefont {E.}~\bibnamefont {Calleja}},
  \bibinfo {author} {\bibfnamefont {J.}~\bibnamefont {Alldredge}}, \bibinfo
  {author} {\bibfnamefont {X.}~\bibnamefont {Zhu}}, \bibinfo {author}
  {\bibfnamefont {L.}~\bibnamefont {Li}}, \bibinfo {author} {\bibfnamefont
  {W.}~\bibnamefont {Lu}}, \bibinfo {author} {\bibfnamefont {Y.}~\bibnamefont
  {Sun}}, \bibinfo {author} {\bibfnamefont {T.}~\bibnamefont {Wolf}}, \bibinfo
  {author} {\bibfnamefont {H.}~\bibnamefont {Berger}}, \ and\ \bibinfo {author}
  {\bibfnamefont {K.}~\bibnamefont {McElroy}},\ }\href@noop {} {\bibfield
  {journal} {\bibinfo  {journal} {Physical Review B}\ }\textbf {\bibinfo
  {volume} {89}},\ \bibinfo {pages} {165140} (\bibinfo {year}
  {2014})}\BibitemShut {NoStop}%
\bibitem [{\citenamefont {Hildebrand}\ \emph {et~al.}(2018)\citenamefont
  {Hildebrand}, \citenamefont {Jaouen}, \citenamefont {Mottas}, \citenamefont
  {Monney}, \citenamefont {Barreteau}, \citenamefont {Giannini}, \citenamefont
  {Bowler},\ and\ \citenamefont {Aebi}}]{Hildebrand2018}%
  \BibitemOpen
  \bibfield  {author} {\bibinfo {author} {\bibfnamefont {B.}~\bibnamefont
  {Hildebrand}}, \bibinfo {author} {\bibfnamefont {T.}~\bibnamefont {Jaouen}},
  \bibinfo {author} {\bibfnamefont {M.~L.}\ \bibnamefont {Mottas}}, \bibinfo
  {author} {\bibfnamefont {G.}~\bibnamefont {Monney}}, \bibinfo {author}
  {\bibfnamefont {C.}~\bibnamefont {Barreteau}}, \bibinfo {author}
  {\bibfnamefont {E.}~\bibnamefont {Giannini}}, \bibinfo {author}
  {\bibfnamefont {D.~R.}\ \bibnamefont {Bowler}}, \ and\ \bibinfo {author}
  {\bibfnamefont {P.}~\bibnamefont {Aebi}},\ }\href {\doibase
  10.1103/PhysRevLett.120.136404} {\bibfield  {journal} {\bibinfo  {journal}
  {Physical Review Letters}\ }\textbf {\bibinfo {volume} {120}},\ \bibinfo
  {pages} {136404} (\bibinfo {year} {2018})}\BibitemShut {NoStop}%
\bibitem [{\citenamefont {Qiao}\ \emph {et~al.}(2017)\citenamefont {Qiao},
  \citenamefont {Zhou}, \citenamefont {Tao}, \citenamefont {Zheng},
  \citenamefont {Wu}, \citenamefont {Ciocys}, \citenamefont {Iavarone},
  \citenamefont {Srolovitz}, \citenamefont {Karapetrov},\ and\ \citenamefont
  {Zhu}}]{qiao2017anisotropic}%
  \BibitemOpen
  \bibfield  {author} {\bibinfo {author} {\bibfnamefont {Q.}~\bibnamefont
  {Qiao}}, \bibinfo {author} {\bibfnamefont {S.}~\bibnamefont {Zhou}}, \bibinfo
  {author} {\bibfnamefont {J.}~\bibnamefont {Tao}}, \bibinfo {author}
  {\bibfnamefont {J.-C.}\ \bibnamefont {Zheng}}, \bibinfo {author}
  {\bibfnamefont {L.}~\bibnamefont {Wu}}, \bibinfo {author} {\bibfnamefont
  {S.~T.}\ \bibnamefont {Ciocys}}, \bibinfo {author} {\bibfnamefont
  {M.}~\bibnamefont {Iavarone}}, \bibinfo {author} {\bibfnamefont {D.~J.}\
  \bibnamefont {Srolovitz}}, \bibinfo {author} {\bibfnamefont {G.}~\bibnamefont
  {Karapetrov}}, \ and\ \bibinfo {author} {\bibfnamefont {Y.}~\bibnamefont
  {Zhu}},\ }\href@noop {} {\bibfield  {journal} {\bibinfo  {journal} {Physical
  Review Materials}\ }\textbf {\bibinfo {volume} {1}},\ \bibinfo {pages}
  {054002} (\bibinfo {year} {2017})}\BibitemShut {NoStop}%
\bibitem [{\citenamefont {Yankovich}\ \emph {et~al.}(2014)\citenamefont
  {Yankovich}, \citenamefont {Berkels}, \citenamefont {Dahmen}, \citenamefont
  {Binev}, \citenamefont {Sanchez}, \citenamefont {Bradley}, \citenamefont
  {Li}, \citenamefont {Szlufarska},\ and\ \citenamefont
  {Voyles}}]{Yankovich2014}%
  \BibitemOpen
  \bibfield  {author} {\bibinfo {author} {\bibfnamefont {A.~B.}\ \bibnamefont
  {Yankovich}}, \bibinfo {author} {\bibfnamefont {B.}~\bibnamefont {Berkels}},
  \bibinfo {author} {\bibfnamefont {W.}~\bibnamefont {Dahmen}}, \bibinfo
  {author} {\bibfnamefont {P.}~\bibnamefont {Binev}}, \bibinfo {author}
  {\bibfnamefont {S.~I.}\ \bibnamefont {Sanchez}}, \bibinfo {author}
  {\bibfnamefont {S.~a.}\ \bibnamefont {Bradley}}, \bibinfo {author}
  {\bibfnamefont {A.}~\bibnamefont {Li}}, \bibinfo {author} {\bibfnamefont
  {I.}~\bibnamefont {Szlufarska}}, \ and\ \bibinfo {author} {\bibfnamefont
  {P.~M.}\ \bibnamefont {Voyles}},\ }\href {\doibase 10.1038/ncomms5155}
  {\bibfield  {journal} {\bibinfo  {journal} {Nature Communications}\ }\textbf
  {\bibinfo {volume} {5}},\ \bibinfo {pages} {4155} (\bibinfo {year}
  {2014})}\BibitemShut {NoStop}%
\bibitem [{\citenamefont {El~Baggari}\ \emph {et~al.}(2018)\citenamefont
  {El~Baggari}, \citenamefont {Savitzky}, \citenamefont {Admasu}, \citenamefont
  {Kim}, \citenamefont {Cheong}, \citenamefont {Hovden},\ and\ \citenamefont
  {Kourkoutis}}]{ElBaggari2018}%
  \BibitemOpen
  \bibfield  {author} {\bibinfo {author} {\bibfnamefont {I.}~\bibnamefont
  {El~Baggari}}, \bibinfo {author} {\bibfnamefont {B.~H.}\ \bibnamefont
  {Savitzky}}, \bibinfo {author} {\bibfnamefont {A.~S.}\ \bibnamefont
  {Admasu}}, \bibinfo {author} {\bibfnamefont {J.}~\bibnamefont {Kim}},
  \bibinfo {author} {\bibfnamefont {S.-W.}\ \bibnamefont {Cheong}}, \bibinfo
  {author} {\bibfnamefont {R.}~\bibnamefont {Hovden}}, \ and\ \bibinfo {author}
  {\bibfnamefont {L.~F.}\ \bibnamefont {Kourkoutis}},\ }\href {\doibase
  10.1073/pnas.1714901115} {\bibfield  {journal} {\bibinfo  {journal}
  {Proceedings of the National Academy of Sciences}\ }\textbf {\bibinfo
  {volume} {115}},\ \bibinfo {pages} {1445} (\bibinfo {year}
  {2018})}\BibitemShut {NoStop}%
\bibitem [{\citenamefont {Hovden}\ \emph {et~al.}(2016)\citenamefont {Hovden},
  \citenamefont {Tsen}, \citenamefont {Liu}, \citenamefont {Savitzky},
  \citenamefont {{El Baggari}}, \citenamefont {Liu}, \citenamefont {Lu},
  \citenamefont {Sun}, \citenamefont {Kim}, \citenamefont {Pasupathy},\ and\
  \citenamefont {Kourkoutis}}]{Hovden2016}%
  \BibitemOpen
  \bibfield  {author} {\bibinfo {author} {\bibfnamefont {R.}~\bibnamefont
  {Hovden}}, \bibinfo {author} {\bibfnamefont {A.~W.}\ \bibnamefont {Tsen}},
  \bibinfo {author} {\bibfnamefont {P.}~\bibnamefont {Liu}}, \bibinfo {author}
  {\bibfnamefont {B.~H.}\ \bibnamefont {Savitzky}}, \bibinfo {author}
  {\bibfnamefont {I.}~\bibnamefont {{El Baggari}}}, \bibinfo {author}
  {\bibfnamefont {Y.}~\bibnamefont {Liu}}, \bibinfo {author} {\bibfnamefont
  {W.}~\bibnamefont {Lu}}, \bibinfo {author} {\bibfnamefont {Y.}~\bibnamefont
  {Sun}}, \bibinfo {author} {\bibfnamefont {P.}~\bibnamefont {Kim}}, \bibinfo
  {author} {\bibfnamefont {A.~N.}\ \bibnamefont {Pasupathy}}, \ and\ \bibinfo
  {author} {\bibfnamefont {L.~F.}\ \bibnamefont {Kourkoutis}},\ }\href
  {\doibase 10.1073/pnas.1606044113} {\bibfield  {journal} {\bibinfo  {journal}
  {Proceedings of the National Academy of Sciences}\ }\textbf {\bibinfo
  {volume} {113}},\ \bibinfo {pages} {11420} (\bibinfo {year}
  {2016})}\BibitemShut {NoStop}%
\bibitem [{\citenamefont {Lee}\ \emph {et~al.}(2019)\citenamefont {Lee},
  \citenamefont {Goh},\ and\ \citenamefont {Cho}}]{lee2019origin}%
  \BibitemOpen
  \bibfield  {author} {\bibinfo {author} {\bibfnamefont {S.-H.}\ \bibnamefont
  {Lee}}, \bibinfo {author} {\bibfnamefont {J.~S.}\ \bibnamefont {Goh}}, \ and\
  \bibinfo {author} {\bibfnamefont {D.}~\bibnamefont {Cho}},\ }\href@noop {}
  {\bibfield  {journal} {\bibinfo  {journal} {Physical review letters}\
  }\textbf {\bibinfo {volume} {122}},\ \bibinfo {pages} {106404} (\bibinfo
  {year} {2019})}\BibitemShut {NoStop}%
\bibitem [{\citenamefont {Ritschel}\ \emph {et~al.}(2018)\citenamefont
  {Ritschel}, \citenamefont {Berger},\ and\ \citenamefont
  {Geck}}]{Ritschel2018}%
  \BibitemOpen
  \bibfield  {author} {\bibinfo {author} {\bibfnamefont {T.}~\bibnamefont
  {Ritschel}}, \bibinfo {author} {\bibfnamefont {H.}~\bibnamefont {Berger}}, \
  and\ \bibinfo {author} {\bibfnamefont {J.}~\bibnamefont {Geck}},\ }\href
  {\doibase 10.1103/PhysRevB.98.195134} {\bibfield  {journal} {\bibinfo
  {journal} {Physical Review B}\ }\textbf {\bibinfo {volume} {98}},\ \bibinfo
  {pages} {195134} (\bibinfo {year} {2018})}\BibitemShut {NoStop}%
\bibitem [{Note1()}]{Note1}%
  \BibitemOpen
  \bibinfo {note} {See Supplemental Material at [URL] for experimental methods,
  additional cross-sectional data, intensity line profiles, details about the
  DFT structural optimization, and multislice simulation parameters and
  results, which includes Refs. \cite
  {Savitzky2018,Momma2008,Campbell2006,Larsen2017,Allen2015,isotrophy}}\BibitemShut
  {NoStop}%
\bibitem [{\citenamefont {Stiehl}\ \emph {et~al.}(2019)\citenamefont {Stiehl},
  \citenamefont {MacNeill}, \citenamefont {Sivadas}, \citenamefont
  {El~Baggari}, \citenamefont {Guimaraes}, \citenamefont {Reynolds},
  \citenamefont {Kourkoutis}, \citenamefont {Fennie}, \citenamefont {Buhrman},\
  and\ \citenamefont {Ralph}}]{stiehl2019current}%
  \BibitemOpen
  \bibfield  {author} {\bibinfo {author} {\bibfnamefont {G.~M.}\ \bibnamefont
  {Stiehl}}, \bibinfo {author} {\bibfnamefont {D.}~\bibnamefont {MacNeill}},
  \bibinfo {author} {\bibfnamefont {N.}~\bibnamefont {Sivadas}}, \bibinfo
  {author} {\bibfnamefont {I.}~\bibnamefont {El~Baggari}}, \bibinfo {author}
  {\bibfnamefont {M.~H.}\ \bibnamefont {Guimaraes}}, \bibinfo {author}
  {\bibfnamefont {N.~D.}\ \bibnamefont {Reynolds}}, \bibinfo {author}
  {\bibfnamefont {L.~F.}\ \bibnamefont {Kourkoutis}}, \bibinfo {author}
  {\bibfnamefont {C.~J.}\ \bibnamefont {Fennie}}, \bibinfo {author}
  {\bibfnamefont {R.~A.}\ \bibnamefont {Buhrman}}, \ and\ \bibinfo {author}
  {\bibfnamefont {D.~C.}\ \bibnamefont {Ralph}},\ }\href@noop {} {\bibfield
  {journal} {\bibinfo  {journal} {ACS nano}\ }\textbf {\bibinfo {volume}
  {13}},\ \bibinfo {pages} {2599} (\bibinfo {year} {2019})}\BibitemShut
  {NoStop}%
\bibitem [{\citenamefont {Fitting}\ \emph {et~al.}(2006)\citenamefont
  {Fitting}, \citenamefont {Thiel}, \citenamefont {Schmehl}, \citenamefont
  {Mannhart},\ and\ \citenamefont {Muller}}]{Fitting2006}%
  \BibitemOpen
  \bibfield  {author} {\bibinfo {author} {\bibfnamefont {L.}~\bibnamefont
  {Fitting}}, \bibinfo {author} {\bibfnamefont {S.}~\bibnamefont {Thiel}},
  \bibinfo {author} {\bibfnamefont {A.}~\bibnamefont {Schmehl}}, \bibinfo
  {author} {\bibfnamefont {J.}~\bibnamefont {Mannhart}}, \ and\ \bibinfo
  {author} {\bibfnamefont {D.~A.}\ \bibnamefont {Muller}},\ }\href {\doibase
  10.1016/j.ultramic.2006.04.019} {\bibfield  {journal} {\bibinfo  {journal}
  {Ultramicroscopy}\ }\textbf {\bibinfo {volume} {106}},\ \bibinfo {pages}
  {1053} (\bibinfo {year} {2006})}\BibitemShut {NoStop}%
\bibitem [{Note2()}]{Note2}%
  \BibitemOpen
  \bibinfo {note} {The bottom layers in contact with silicon oxide are exposed
  to air and show defects and amorphous features. The top layer which is
  exposed under high vaccuum and capped with permalloy maintains the trimerized
  distortion \cite {Note1,stiehl2019current}.}\BibitemShut {Stop}%
\bibitem [{Note3()}]{Note3}%
  \BibitemOpen
  \bibinfo {note} {Using convergent beam electron diffraction pattern, we
  estimate the thickness of the specimen to be between 18 and 22 nm (27-33
  layers).}\BibitemShut {Stop}%
\bibitem [{\citenamefont {Perovic}\ \emph {et~al.}(1993)\citenamefont
  {Perovic}, \citenamefont {Rossouw},\ and\ \citenamefont
  {Howie}}]{perovic1993imaging}%
  \BibitemOpen
  \bibfield  {author} {\bibinfo {author} {\bibfnamefont {D.}~\bibnamefont
  {Perovic}}, \bibinfo {author} {\bibfnamefont {C.}~\bibnamefont {Rossouw}}, \
  and\ \bibinfo {author} {\bibfnamefont {A.}~\bibnamefont {Howie}},\
  }\href@noop {} {\bibfield  {journal} {\bibinfo  {journal} {Ultramicroscopy}\
  }\textbf {\bibinfo {volume} {52}},\ \bibinfo {pages} {353} (\bibinfo {year}
  {1993})}\BibitemShut {NoStop}%
\bibitem [{\citenamefont {Hillyard}\ and\ \citenamefont
  {Silcox}(1995)}]{hillyard1995detector}%
  \BibitemOpen
  \bibfield  {author} {\bibinfo {author} {\bibfnamefont {S.}~\bibnamefont
  {Hillyard}}\ and\ \bibinfo {author} {\bibfnamefont {J.}~\bibnamefont
  {Silcox}},\ }\href@noop {} {\bibfield  {journal} {\bibinfo  {journal}
  {Ultramicroscopy}\ }\textbf {\bibinfo {volume} {58}},\ \bibinfo {pages} {6}
  (\bibinfo {year} {1995})}\BibitemShut {NoStop}%
\bibitem [{\citenamefont {Haruta}\ \emph {et~al.}(2009)\citenamefont {Haruta},
  \citenamefont {Kurata}, \citenamefont {Komatsu}, \citenamefont {Shimakawa},\
  and\ \citenamefont {Isoda}}]{Haruta2009}%
  \BibitemOpen
  \bibfield  {author} {\bibinfo {author} {\bibfnamefont {M.}~\bibnamefont
  {Haruta}}, \bibinfo {author} {\bibfnamefont {H.}~\bibnamefont {Kurata}},
  \bibinfo {author} {\bibfnamefont {H.}~\bibnamefont {Komatsu}}, \bibinfo
  {author} {\bibfnamefont {Y.}~\bibnamefont {Shimakawa}}, \ and\ \bibinfo
  {author} {\bibfnamefont {S.}~\bibnamefont {Isoda}},\ }\href {\doibase
  10.1016/j.ultramic.2009.01.004} {\bibfield  {journal} {\bibinfo  {journal}
  {Ultramicroscopy}\ }\textbf {\bibinfo {volume} {109}},\ \bibinfo {pages}
  {361} (\bibinfo {year} {2009})}\BibitemShut {NoStop}%
\bibitem [{\citenamefont {Esser}\ \emph {et~al.}(2016)\citenamefont {Esser},
  \citenamefont {Hauser}, \citenamefont {Williams}, \citenamefont {Allen},
  \citenamefont {Woodward}, \citenamefont {Yang},\ and\ \citenamefont
  {McComb}}]{Esser2016}%
  \BibitemOpen
  \bibfield  {author} {\bibinfo {author} {\bibfnamefont {B.~D.}\ \bibnamefont
  {Esser}}, \bibinfo {author} {\bibfnamefont {A.~J.}\ \bibnamefont {Hauser}},
  \bibinfo {author} {\bibfnamefont {R.~E.~A.}\ \bibnamefont {Williams}},
  \bibinfo {author} {\bibfnamefont {L.~J.}\ \bibnamefont {Allen}}, \bibinfo
  {author} {\bibfnamefont {P.~M.}\ \bibnamefont {Woodward}}, \bibinfo {author}
  {\bibfnamefont {F.~Y.}\ \bibnamefont {Yang}}, \ and\ \bibinfo {author}
  {\bibfnamefont {D.~W.}\ \bibnamefont {McComb}},\ }\href@noop {} {\bibfield
  {journal} {\bibinfo  {journal} {Physical review letters}\ }\textbf {\bibinfo
  {volume} {117}},\ \bibinfo {pages} {176101} (\bibinfo {year}
  {2016})}\BibitemShut {NoStop}%
\bibitem [{\citenamefont {Savitzky}\ \emph {et~al.}(2017)\citenamefont
  {Savitzky}, \citenamefont {El~Baggari}, \citenamefont {Admasu}, \citenamefont
  {Kim}, \citenamefont {Cheong}, \citenamefont {Hovden},\ and\ \citenamefont
  {Kourkoutis}}]{savitzky2017bending}%
  \BibitemOpen
  \bibfield  {author} {\bibinfo {author} {\bibfnamefont {B.~H.}\ \bibnamefont
  {Savitzky}}, \bibinfo {author} {\bibfnamefont {I.}~\bibnamefont
  {El~Baggari}}, \bibinfo {author} {\bibfnamefont {A.~S.}\ \bibnamefont
  {Admasu}}, \bibinfo {author} {\bibfnamefont {J.}~\bibnamefont {Kim}},
  \bibinfo {author} {\bibfnamefont {S.-W.}\ \bibnamefont {Cheong}}, \bibinfo
  {author} {\bibfnamefont {R.}~\bibnamefont {Hovden}}, \ and\ \bibinfo {author}
  {\bibfnamefont {L.~F.}\ \bibnamefont {Kourkoutis}},\ }\href@noop {}
  {\bibfield  {journal} {\bibinfo  {journal} {Nature communications}\ }\textbf
  {\bibinfo {volume} {8}},\ \bibinfo {pages} {1883} (\bibinfo {year}
  {2017})}\BibitemShut {NoStop}%
\bibitem [{Note4()}]{Note4}%
  \BibitemOpen
  \bibinfo {note} {Extracting a reference image from the simulated data yields
  the same dominant intensity modulations associated with the trimer
  distortion, albeit with some minor differences in the background intensity
  \cite {Note1}.}\BibitemShut {Stop}%
\bibitem [{\citenamefont {Kresse}\ and\ \citenamefont
  {Furthm{\"{u}}ller}(1996)}]{Kresse1996}%
  \BibitemOpen
  \bibfield  {author} {\bibinfo {author} {\bibfnamefont {G.}~\bibnamefont
  {Kresse}}\ and\ \bibinfo {author} {\bibfnamefont {J.}~\bibnamefont
  {Furthm{\"{u}}ller}},\ }\href {\doibase 10.1103/PhysRevB.54.11169} {\bibfield
   {journal} {\bibinfo  {journal} {Physical Review B}\ }\textbf {\bibinfo
  {volume} {54}},\ \bibinfo {pages} {11169} (\bibinfo {year}
  {1996})}\BibitemShut {NoStop}%
\bibitem [{\citenamefont {Kresse}\ and\ \citenamefont
  {Joubert}(1999)}]{Kresse1999}%
  \BibitemOpen
  \bibfield  {author} {\bibinfo {author} {\bibfnamefont {G.}~\bibnamefont
  {Kresse}}\ and\ \bibinfo {author} {\bibfnamefont {D.}~\bibnamefont
  {Joubert}},\ }\href {\doibase 10.1103/PhysRevB.59.1758} {\bibfield  {journal}
  {\bibinfo  {journal} {Physical Review B}\ }\textbf {\bibinfo {volume} {59}},\
  \bibinfo {pages} {1758} (\bibinfo {year} {1999})}\BibitemShut {NoStop}%
\bibitem [{\citenamefont {Perdew}\ \emph {et~al.}(2008)\citenamefont {Perdew},
  \citenamefont {Ruzsinszky}, \citenamefont {Csonka}, \citenamefont {Vydrov},
  \citenamefont {Scuseria}, \citenamefont {Constantin}, \citenamefont {Zhou},\
  and\ \citenamefont {Burke}}]{perdew2008restoring}%
  \BibitemOpen
  \bibfield  {author} {\bibinfo {author} {\bibfnamefont {J.~P.}\ \bibnamefont
  {Perdew}}, \bibinfo {author} {\bibfnamefont {A.}~\bibnamefont {Ruzsinszky}},
  \bibinfo {author} {\bibfnamefont {G.~I.}\ \bibnamefont {Csonka}}, \bibinfo
  {author} {\bibfnamefont {O.~A.}\ \bibnamefont {Vydrov}}, \bibinfo {author}
  {\bibfnamefont {G.~E.}\ \bibnamefont {Scuseria}}, \bibinfo {author}
  {\bibfnamefont {L.~A.}\ \bibnamefont {Constantin}}, \bibinfo {author}
  {\bibfnamefont {X.}~\bibnamefont {Zhou}}, \ and\ \bibinfo {author}
  {\bibfnamefont {K.}~\bibnamefont {Burke}},\ }\href@noop {} {\bibfield
  {journal} {\bibinfo  {journal} {Physical review letters}\ }\textbf {\bibinfo
  {volume} {100}},\ \bibinfo {pages} {136406} (\bibinfo {year}
  {2008})}\BibitemShut {NoStop}%
\bibitem [{\citenamefont {Savitzky}\ \emph {et~al.}(2018)\citenamefont
  {Savitzky}, \citenamefont {{El Baggari}}, \citenamefont {Clement},
  \citenamefont {Waite}, \citenamefont {Goodge}, \citenamefont {Baek},
  \citenamefont {Sheckelton}, \citenamefont {Pasco}, \citenamefont {Nair},
  \citenamefont {Schreiber}, \citenamefont {Hoffman}, \citenamefont {Admasu},
  \citenamefont {Kim}, \citenamefont {Cheong}, \citenamefont {Bhattacharya},
  \citenamefont {Schlom}, \citenamefont {McQueen}, \citenamefont {Hovden},\
  and\ \citenamefont {Kourkoutis}}]{Savitzky2018}%
  \BibitemOpen
  \bibfield  {author} {\bibinfo {author} {\bibfnamefont {B.~H.}\ \bibnamefont
  {Savitzky}}, \bibinfo {author} {\bibfnamefont {I.}~\bibnamefont {{El
  Baggari}}}, \bibinfo {author} {\bibfnamefont {C.~B.}\ \bibnamefont
  {Clement}}, \bibinfo {author} {\bibfnamefont {E.}~\bibnamefont {Waite}},
  \bibinfo {author} {\bibfnamefont {B.~H.}\ \bibnamefont {Goodge}}, \bibinfo
  {author} {\bibfnamefont {D.~J.}\ \bibnamefont {Baek}}, \bibinfo {author}
  {\bibfnamefont {J.~P.}\ \bibnamefont {Sheckelton}}, \bibinfo {author}
  {\bibfnamefont {C.}~\bibnamefont {Pasco}}, \bibinfo {author} {\bibfnamefont
  {H.}~\bibnamefont {Nair}}, \bibinfo {author} {\bibfnamefont {N.~J.}\
  \bibnamefont {Schreiber}}, \bibinfo {author} {\bibfnamefont {J.}~\bibnamefont
  {Hoffman}}, \bibinfo {author} {\bibfnamefont {A.~S.}\ \bibnamefont {Admasu}},
  \bibinfo {author} {\bibfnamefont {J.}~\bibnamefont {Kim}}, \bibinfo {author}
  {\bibfnamefont {S.~W.}\ \bibnamefont {Cheong}}, \bibinfo {author}
  {\bibfnamefont {A.}~\bibnamefont {Bhattacharya}}, \bibinfo {author}
  {\bibfnamefont {D.~G.}\ \bibnamefont {Schlom}}, \bibinfo {author}
  {\bibfnamefont {T.~M.}\ \bibnamefont {McQueen}}, \bibinfo {author}
  {\bibfnamefont {R.}~\bibnamefont {Hovden}}, \ and\ \bibinfo {author}
  {\bibfnamefont {L.~F.}\ \bibnamefont {Kourkoutis}},\ }\href {\doibase
  10.1016/j.ultramic.2018.04.008} {\bibfield  {journal} {\bibinfo  {journal}
  {Ultramicroscopy}\ }\textbf {\bibinfo {volume} {191}} (\bibinfo {year}
  {2018}),\ 10.1016/j.ultramic.2018.04.008}\BibitemShut {NoStop}%
\bibitem [{\citenamefont {Momma}\ and\ \citenamefont
  {Izumi}(2008)}]{Momma2008}%
  \BibitemOpen
  \bibfield  {author} {\bibinfo {author} {\bibfnamefont {K.}~\bibnamefont
  {Momma}}\ and\ \bibinfo {author} {\bibfnamefont {F.}~\bibnamefont {Izumi}},\
  }\href {\doibase 10.1107/S0021889808012016} {\bibfield  {journal} {\bibinfo
  {journal} {Journal of Applied Crystallography}\ }\textbf {\bibinfo {volume}
  {41}},\ \bibinfo {pages} {653} (\bibinfo {year} {2008})}\BibitemShut
  {NoStop}%
\bibitem [{\citenamefont {Campbell}\ \emph {et~al.}(2006)\citenamefont
  {Campbell}, \citenamefont {Stokes}, \citenamefont {Tanner},\ and\
  \citenamefont {Hatch}}]{Campbell2006}%
  \BibitemOpen
  \bibfield  {author} {\bibinfo {author} {\bibfnamefont {B.~J.}\ \bibnamefont
  {Campbell}}, \bibinfo {author} {\bibfnamefont {H.~T.}\ \bibnamefont
  {Stokes}}, \bibinfo {author} {\bibfnamefont {D.~E.}\ \bibnamefont {Tanner}},
  \ and\ \bibinfo {author} {\bibfnamefont {D.~M.}\ \bibnamefont {Hatch}},\
  }\href {\doibase 10.1107/S0021889806014075} {\bibfield  {journal} {\bibinfo
  {journal} {Journal of Applied Crystallography}\ }\textbf {\bibinfo {volume}
  {39}},\ \bibinfo {pages} {607} (\bibinfo {year} {2006})}\BibitemShut
  {NoStop}%
\bibitem [{\citenamefont {Larsen}\ \emph {et~al.}(2017)\citenamefont {Larsen},
  \citenamefont {Jens}, \citenamefont {Jakob}, \citenamefont {Ivano},
  \citenamefont {Rune}, \citenamefont {Marcin}, \citenamefont {Jesper},
  \citenamefont {Michael}, \citenamefont {Bjork},\ and\ \citenamefont
  {Cory}}]{Larsen2017}%
  \BibitemOpen
  \bibfield  {author} {\bibinfo {author} {\bibfnamefont {A.}~\bibnamefont
  {Larsen}}, \bibinfo {author} {\bibfnamefont {M.}~\bibnamefont {Jens}},
  \bibinfo {author} {\bibfnamefont {B.}~\bibnamefont {Jakob}}, \bibinfo
  {author} {\bibfnamefont {C.}~\bibnamefont {Ivano}}, \bibinfo {author}
  {\bibfnamefont {C.}~\bibnamefont {Rune}}, \bibinfo {author} {\bibfnamefont
  {D.}~\bibnamefont {Marcin}}, \bibinfo {author} {\bibfnamefont
  {F.}~\bibnamefont {Jesper}}, \bibinfo {author} {\bibfnamefont
  {G.}~\bibnamefont {Michael}}, \bibinfo {author} {\bibfnamefont
  {H.}~\bibnamefont {Bjork}}, \ and\ \bibinfo {author} {\bibfnamefont
  {H.}~\bibnamefont {Cory}},\ }\href {\doibase 10.1088/1361-648X/aa680e}
  {\bibfield  {journal} {\bibinfo  {journal} {Journal of Physics Condensed
  Matter}\ }\textbf {\bibinfo {volume} {29}},\ \bibinfo {pages} {27} (\bibinfo
  {year} {2017})}\BibitemShut {NoStop}%
\bibitem [{\citenamefont {Allen}\ \emph {et~al.}(2015)\citenamefont {Allen},
  \citenamefont {D'Alfonso},\ and\ \citenamefont {Findlay}}]{Allen2015}%
  \BibitemOpen
  \bibfield  {author} {\bibinfo {author} {\bibfnamefont {L.~J.}\ \bibnamefont
  {Allen}}, \bibinfo {author} {\bibfnamefont {A.~J.}\ \bibnamefont
  {D'Alfonso}}, \ and\ \bibinfo {author} {\bibfnamefont {S.~D.}\ \bibnamefont
  {Findlay}},\ }\href {\doibase 10.1016/j.ultramic.2014.10.011} {\bibfield
  {journal} {\bibinfo  {journal} {Ultramicroscopy}\ }\textbf {\bibinfo {volume}
  {151}},\ \bibinfo {pages} {11} (\bibinfo {year} {2015})}\BibitemShut
  {NoStop}%
\bibitem [{\citenamefont {Stokes}\ \emph {et~al.}()\citenamefont {Stokes},
  \citenamefont {Hatch},\ and\ \citenamefont {Campbell}}]{isotrophy}%
  \BibitemOpen
  \bibfield  {author} {\bibinfo {author} {\bibfnamefont {H.~T.}\ \bibnamefont
  {Stokes}}, \bibinfo {author} {\bibfnamefont {D.~M.}\ \bibnamefont {Hatch}}, \
  and\ \bibinfo {author} {\bibfnamefont {B.~J.}\ \bibnamefont {Campbell}},\
  }\href@noop {} {\enquote {\bibinfo {title} {{ISOTROPY Software Suite}},}\
  }\bibinfo {howpublished} {\url{iso.byu.edu}}\BibitemShut {NoStop}%
\end{thebibliography}%
\end{document}